\documentstyle[12pt]{article}
\begin{document}

\title{Experimental constraints on the parameter space of the
next-to-minimal supersymmetric standard model at LEP 2
       \thanks{This work is supported in part by Ministry of Education, 
        Korea, BSRI-97-2442.}}
\author{S.W. Ham$^1$, S.K. Oh$^1$, and B.R. Kim$^2$
\\
\\
        $^1$ Department of Physics, Kon-Kuk University,\\
         Seoul 143-701, Korea
\\
        $^2$ III. Physikalisches. Institut. A, RWTH Aachen, \\
        D52056 Aachen, Germany}
\date{}
\maketitle
\begin{abstract}
We search for the neutral Higgs sector of the next-to-minimal supersymmetric
standard model at LEP 2.
At the tree level any experimental constraints on $\tan \beta$ cannot be set
by the Higgs search at LEP 2 with $\sqrt{s}$ = 175 GeV, whereas
at LEP 2 with $\sqrt{s}$ = 192 GeV $\tan \beta$ can be set by an experimental
constraint.
Furthermore the tree level parameter space of the model can be completely
explored by the Higgs search at LEP 2 with $\sqrt{s}$ = 205 GeV.
Radiative corrections both to the neutral Higgs boson masses and to the
relevant couplings for the scalar Higgs productions give large contributions
to the production cross sections of the scalar Higgs bosons at the tree level.
The tree level situation at LEP 2 with $\sqrt{s}$ = 192 GeV as well as
with $\sqrt{s}$ = 205 GeV can be drastically changed by these effects.
We expect that a small region of the 1-loop level parameter space of the model
via the scalar Higgs production can be explored by the Higgs search at LEP 2.
\end{abstract}
\vfil

%***********************************************************************
\section{Introduction}
%***********************************************************************

\hspace*{5.5mm}
Many theoretical particle physicists believe that the sucess of the standard
model (SM) at the electroweak scale will not last in higher energy scales.
This belief arises from several defects of the SM, such as, the gauge hiearchy
problem and the naturalness problem.
Supersymmetry (SUSY) has been proposed to resolve these problems in a
technical way.
The particle content in the SUSY models consists of the SM particles and their
superpartners.
The SM particles and their superpartners have equal masses but different spins.
Thus quadratic divergence due to the 1-loop corrections to the Higgs boson mass
would be absent in the SUSY models because the SM particle and its superpartner
loops have opposite signs.
However, since SUSY is a broken symmetry in nature, the tree level mass of the
Higgs boson receives non-zero quadratic corrections due to an incomplete
cancellation between the SM particle and its superpartner loops.

In supersymmetric models, at least two Higgs doublets ($H_1$, $H_2$) are
required to give masses to fermions via spontaneous symmetry breaking.
The masses of down-type fermions are generated in terms of the vacuum
expectation value ($v_1$) of $H_1$ while the masses of up-type fermions
are generated in terms of the vacuum expectation value ($v_2$) of $H_2$.
Among various non-minimal supersymmetric models, the simplest one is the
next-to-minimal supersymmetric standard model (NMSSM).
Its Higgs sector consists of two Higgs doublets $H_1$ and $H_2$ and one Higgs
singlet $N$.
Thus the model has ten degrees of freedom in the Higgs sector.
After spontaneous symmetry breaking, three neutral scalar Higgs bosons
$(S_1, S_2, S_3)$, two neutral pseudoscalar Higgs bosons $(P_1, P_2)$,
and a pair of charged Higgs bosons $(H^{\pm})$ emerge as physical particles.

Recently researches for the scalar Higgs boson mass in the NMSSM using the
effective potential method have been performed by several authors.
It is well known that radiative corrections due to the top-quark and
top-scalar-quark loops give large contributions to the tree level mass of
the scalar Higgs boson.
The radiative corrections calculated so far may be classified as follows:
(i) assuming the degeneracy of the left- and right-handed top-scalar-quark
masses [1], (ii) without the degeneracy assumption of the left- and
right-handed top-scalar-quark masses [2], (iii) with the non-degeneracy
assumption of the left- and right-handed top-scalar-quark masses as well as
with the partial gauge boson contributions [3].
Note that both (i) and (ii) did not include terms proportional to gauge
couplings in the top-scalar-quark masses [3].
According to Ham et al. [3], the absolute upper bound on the lightest scalar
Higgs boson mass including the partial gauge boson contributions at the 1-loop
level is about 156 GeV for a reasonable parameter space.

In $e^+e^-$ collsions, there are three dominant processes for the scalar Higgs
production.
Three processes are known as the Higgs-strahlung process, the bremsstrahlung
process, and the associated pair production process.
Useful formuas for the production cross section of the scalar Higgs boson
in the NMSSM are present in the previous papers [4].
Kim et al. [5] remark that LEP 1 data do not exclude the existence of a
massless scalar Higgs boson in the NMSSM at the tree level.
Franke et al. [6] obtained the conclusion that small values of singlet
vacuum expectation in the NMSSM parameter space are excluded by the negative
experiment results of the Higgs and neutralino search at LEP 1.

In this paper we search for the neutral Higgs sector of the NMSSM at LEP 2.
We calculate the lower bounds on the scalar Higgs boson masses at the tree
level as well as at the 1-loop level.
These bounds at the 1-loop level partially contain the gauge boson
contributions, since the top-scalar-quark masses in the 1-loop effective
potential include terms proportional to gauge couplings.
Although three scalar Higgs productions both at the tree level and at the
1-loop level are kinematically allowed at the proposed center of mass
energies of 175, 192, and 205 GeV for LEP 2, at both levels two $S_1$ and $S_2$
can be dominantly produced at these c.m. energies.
We calculate the combined production cross section of $S_1$ and $S_2$ at the
tree level.
Then we impose experimental constraints on $\tan \beta$ using the discovery
potential of LEP 2.
The production cross sections of the scalar Higgs bosons at the tree level can
be largely affected by radiative corrections both to the neutral Higgs boson
masses and to the relevant couplings for the scalar productions.
We calculate the combined production cross section of $S_1$ and $S_2$
including these contributions at three c.m. energies for LEP 2.
Then we investigate how much fraction of the 1-loop level parameter space of
the model can be explored by the Higgs search at LEP 2.

%***********************************************************************
\section{The neutral Higgs sector}
%***********************************************************************

\hspace*{5.5mm}
In the NMSSM, one Higgs singlet $N$ is a neutral field.
Two Higgs doublets are defined as
\begin{eqnarray}
\begin{array}{ccccc}
H_1 & = & \left ( \begin{array}{c}
          H_1^0   \cr
          H_1^-
  \end{array} \right )
\ \ \ \ , \ \ \ \
H_2 & = & \left ( \begin{array}{c}
          H_2^+               \cr
          H_2^0
  \end{array} \right )
\end{array}           \ .
\end{eqnarray}
The tree level Higgs potential of the NMSSM can be decomposed as the neutral
parts and the charged parts with respect to the Higgs fields.
The neutral Higgs potential at the tree level is given as
\begin{eqnarray}
   V_{\rm tree} & = & |\lambda|^2[(|H_1^0|^2+|H_2^0|^2)|N|^2+|H_1^0 H_2^0|^2]
              + |k|^2|N|^4 - (\lambda k^*H_1^0 H_2^0 N^{*2}+{\rm h.c.})  \cr
       & &\mbox{}+{g_1^2 + g_2^2 \over 8}
              (|H_1^0|^2 - |H_2^0|^2)^2 + V_{\rm soft } \ ,
\end{eqnarray}
with
\begin{eqnarray}
     V_{\rm soft} & = & m_{H_1}^2 |H_1^0|^2 + m_{H_2}^2 |H_2^0|^2
                        +m_N^2 |N|^2
        -(\lambda A_\lambda H_1^0 H_2^0 N + {1\over 3} k A_k N^3
        + {\rm h.c.}) \ ,
\end{eqnarray}
where $g_1$ and $g_2$ are the U(1) and SU(2) gauge coupling constants,
respectively.
$A_k$ and $A_\lambda$ are the trilinear soft SUSY breaking parameters with
mass dimension.
$m_{H_1}$, $m_{H_2}$, and $m_N$ are the soft SUSY breaking parameters.
We assume $\lambda$ and $k$ as real and positive parameters.

Now let us consider radiative corrections to the neutral Higgs boson mass.
It is well known that radiative corrections due to the top-quark and
top-scalar-quark loops give a large contribution to the tree level mass of
the lightest scalar Higgs boson.
Therefore we consider radiative corrections to the neutral Higgs boson
mass due to the top-quark and top-scalar-quark loops.
Here we employ the effective potential method in order to calculate radiative
corrections to the tree level mass of the neutral Higgs boson.
The 1-loop effective potential due to the top-quark and top-scalar-quark loops
is expressed as [2,3]
\begin{eqnarray}
   V_{\rm 1-loop} & = &\frac{3}{32\pi^2} {\cal M}_{\tilde{t_i}}^4
   \left (\log {{\cal M}_{\tilde{t_i}}^2 \over \Lambda^2} - {3\over 2} \right )
   - \frac{3}{16\pi^2} {\cal M}_t^4
        \left (\log {{\cal M}_t^2 \over \Lambda^2}  - {3\over 2} \right )  \ ,
\end{eqnarray}
where the mass-squared of the top-quark depending on the neutal Higgs fields
is given by
\begin{eqnarray}
    {\cal M}_t^2 = h_t^2 |H_2^0|^2
\end{eqnarray}
and the mass-squareds of the top-scalar-quarks depending on the neutral Higgs
fields are given by
\begin{eqnarray}
     {\cal M}_{\tilde{t_1}, \tilde{t_2}}^2 & = &
     h_t^2 |H_2^0|^2 + {1 \over 2}(m_Q^2 + m_T^2)
     + {1 \over 8}(g_1^2 + g_2^2)(|H_1^0|^2 - |H_2^0|^2)  \\
     & &\mbox{}\mp \sqrt{ \left [ {1 \over 2}(m_Q^2 - m_T^2)
     +({1 \over 4} g_2^2 - {5 \over 12} g_1^2)
     (|H_1^0|^2 - |H_2^0|^2) \right ]^2
     + h_t^2 |A_t H_2^{0*} + \lambda N H_1^0 |^2}  \ . \nonumber
\end{eqnarray}
In the above equations $h_t$ is the top-quark Yukawa coupling, $A_t$ is the
trilinear soft SUSY breaking parameter, and $m_Q$ and $m_T$ are the soft SUSY
breaking scalar-quark masses.
The arbitrary scale $\Lambda$ is taken to be $M_{\rm SUSY}$ = 1 TeV.
Then the mass-squareds of the top-quark and top-scalar-quarks are given as
\begin{eqnarray}
  m_t^2 & = & {\cal M}_t^2 | _{<H^0_1> = v_1,\, <H^0_2> = v_2,\, <N> = x}  \cr
  m_{\tilde{t_1},\, \tilde{t_2}}^2 & = & {\cal M}_{\tilde{t_1}, \tilde{t_2}}^2
        | _{<H^0_1> = v_1,\, <H^0_2> = v_2,\, <N> = x} \ ,
\end{eqnarray}
$m_{\tilde{t_1}}^2 < m_{\tilde{t_2}}^2$.

The full scalar Higgs potential up to the 1-loop level is given as
\begin{eqnarray}
V = V_{\rm tree} + V_{\rm 1-loop} \ .
\end{eqnarray}
The elements of the mass-squared matrix for the neutral Higgs boson are given by
\begin{equation}
        M_{ij} = \left. \left(
        {\partial^2 V \over \partial \phi_i \partial \phi_j}
        \right)
        \right| _{<H^0_1> = v_1,\, <H^0_2> = v_2,\, <N> = x} \ ,
\end{equation}
where $\phi_i$ are the conventional notations for the real and imaginary
parts of the Higgs fields.
The conditions that the Higgs potential $V$ becomes minimum at
$<H_1^0> = v_1$, $<H_2^0> = v_2$, and $<N>= x$ yield three constraints,
which can eliminate the soft SUSY breaking parameters $m_{H_1}$, $m_{H_2}$,
and $m_N$ from the potential $V$.
Then the scalar and pseudoscalar Higgs boson mass-squared matrices depend on
$\Lambda$, $m_t$, $m_Q$, $m_T$, $A_t$ as the 1-loop level parameters as well
as $k$, $\tan \beta$, $\lambda$, $A_{\lambda}$, $A_k$, $x$ as the tree level
parameters.

The complex formula for the scalar Higgs boson mass-squared matrix $M^S$
is presented in the previous paper [3].
Here we simply write down the elements of the symmetric $3 \times 3$
mass-squared matrix for the scalar Higgs boson.
Their elements $M_{S_{ij}}$ are
\begin{eqnarray}
        M_{S_{11}} & = & (m_Z \cos \beta)^2
            + \lambda x \tan \beta (A_{\lambda} + kx)  \cr
        & &\mbox{} + {3 \over 8 \pi^2} \Delta_1^2
        g(m_{\tilde{t_1}}^2,m_{\tilde{t_2}}^2)
        +{3 m_Z^4 \cos^2 \beta \over 128 \pi^2 v^2}
        \log {m_{\tilde{t_1}}^2 m_{\tilde{t_2}}^2\over \Lambda^4}  \cr
        & &\mbox{} + {3 \over 16 \pi^2 v^2}
        \left [ {2 m_t^2 A_t \lambda x \over \sin 2 \beta}
        - ({4 \over 3} m_W^2 - {5 \over 6} m_Z^2)^2 \cos^2 \beta \right ]
        f(m_{\tilde{t_1}}^2,m_{\tilde{t_2}}^2)       \cr
        & &\mbox{} + {3 \over 16 \pi^2 v} m_Z^2 \cos \beta
  \left ({\Delta_1 \over m_{\tilde{t_1}}^2 - m_{\tilde{t_2}}^2} \right )
        \log {m_{\tilde{t_1}}^2 \over m_{\tilde{t_2}}^2}   \ ,\cr
        & &  \cr
        M_{S_{22}} & = & (m_Z \sin \beta)^2
            + \lambda x \cot \beta (A_{\lambda} + kx)  \cr
        & &\mbox{}+ {3 \over 8 \pi^2} \Delta_2^2
        g(m_{\tilde{t_1}}^2,m_{\tilde{t_2}}^2)
        - {3 m_t^4 \over 4 \pi^2 v^2 \sin^2 \beta}
        \log {m_t^2 \over \Lambda^2}  \cr
        & &\mbox{} + {3 \over 16 \pi^2 v^2}
        \left [ {m_t^2 A_t \lambda x \cot \beta \over \sin^2 \beta}
        - ({4 \over 3} m_W^2 - {5 \over 6} m_Z^2)^2 \sin^2 \beta \right ]
        f(m_{\tilde{t_1}}^2,m_{\tilde{t_2}}^2)       \cr
        & &\mbox{} + {3 \over 16 \pi^2 v}
        \left ({4 m_t^2 \over \sin \beta} - m_Z^2 \sin \beta \right )
   \left ({\Delta_2 \over m_{\tilde{t_1}}^2 - m_{\tilde{t_2}}^2} \right )
        \log {m_{\tilde{t_1}}^2 \over m_{\tilde{t_2}}^2}  \cr
        & &\mbox{} + {3 \over 32 \pi^2 v^2}
        \left ({2 m_t^2 \over \sin \beta}
        - {1 \over 2} m_Z^2 \sin \beta \right )^2
        \log {m_{\tilde{t_1}}^2 m_{\tilde{t_2}}^2 \over \Lambda^4}  \ , \cr
        & &  \cr
        M_{S_{33}} & = & (2 k x)^2 - k x A_k
            + {\lambda \over 2x} v^2 A_{\lambda} \sin 2 \beta
        + {3 \over 16 \pi^2 x} m_t^2 A_t \lambda \cot \beta
        f(m_{\tilde{t_1}}^2,m_{\tilde{t_2}}^2)       \cr
        & &\mbox{}+ {3 \over 8 \pi^2} m_t^4 \lambda^2 \cot^2 \beta
        (A_t + \lambda x \cot \beta)^2
        g(m_{\tilde{t_1}}^2,m_{\tilde{t_2}}^2)    \ ,    \cr
        & & \cr
        M_{S_{12}} & = & (\lambda^2 v^2 - {1 \over 2} m_Z^2) \sin 2 \beta
            - \lambda x (A_{\lambda} +kx)     \cr
        & &\mbox{}+ {3 \over 8 \pi^2} \Delta_1 \Delta_2
        g(m_{\tilde{t_1}}^2,m_{\tilde{t_2}}^2)
        + {3 m_Z^2 \sin 2 \beta \over 256 \pi^2 v^2}
        \left( {4 m_t^2 \over \sin^2 \beta } -m_Z^2 \right)
        \log {m_{\tilde{t_1}}^2 m_{\tilde{t_2}}^2 \over \Lambda^4}  \cr
        & &\mbox{} + {3 \over 32 \pi^2 v^2}
        \left [({4 \over 3} m_W^2 - {5 \over 6} m_Z^2)^2 \sin 2 \beta
        - {2 m_t^2 A_t \lambda x \over \sin^2 \beta} \right]
        f(m_{\tilde{t_1}}^2,m_{\tilde{t_2}}^2)       \cr
        & &\mbox{} + {3 \over 32 \pi^2 v} \left [m_Z^2 \cos \beta \Delta_2
     + ({4 m_t^2 \over \sin^2 \beta} - m_Z^2) \sin \beta \Delta_1 \right]
        {1 \over (m_{\tilde{t_1}}^2 - m_{\tilde{t_2}}^2)}
        \log {m_{\tilde{t_1}}^2 \over m_{\tilde{t_2}}^2}  \ , \cr
        & & \cr
        M_{S_{23}} & = & 2 \lambda^2 x v \sin \beta
        - \lambda v \cos \beta (A_{\lambda} + 2 k x)
        - {3 m_t^2 A_t \lambda \cot \beta \over 16 \pi^2 v \sin \beta}
        f(m_{\tilde{t_1}}^2,m_{\tilde{t_2}}^2)       \cr
        & &\mbox{}+ {3 \over 8 \pi^2} \Delta_2 m_t^2 \lambda \cot \beta
        (A_t + \lambda x \cot \beta)
        g(m_{\tilde{t_1}}^2,m_{\tilde{t_2}}^2) \cr
        & &\mbox{} + {3 m_t^2 \lambda \cot \beta \over 32 \pi^2 v}
        \left ({4 m_t^2 \over \sin \beta} - m_Z^2 \sin \beta \right)
        \left ({ A_t + \lambda x \cot \beta
        \over m_{\tilde{t_1}}^2 - m_{\tilde{t_2}}^2} \right)
        \log {m_{\tilde{t_1}}^2 \over m_{\tilde{t_2}}^2}  \ , \cr
        & &  \cr
        M_{S_{13}} & = & 2 \lambda^2 x v \cos \beta
            - \lambda v \sin \beta (A_{\lambda} +2kx) \cr
        & &\mbox{}+ {3 \over 8 \pi^2} \Delta_1 m_t^2 \lambda \cot \beta
        (A_t + \lambda x \cot \beta)
        g(m_{\tilde{t_1}}^2,m_{\tilde{t_2}}^2) \cr
        & &\mbox{}
        + {3 m_Z^2 m_t^2 \lambda \cos \beta \cot \beta \over 32 \pi^2 v}
        \left ({ A_t + \lambda x \cot \beta
        \over m_{\tilde{t_1}}^2 - m_{\tilde{t_2}}^2} \right)
        \log {m_{\tilde{t_1}}^2 \over m_{\tilde{t_2}}^2}  \cr
        & &\mbox{}
        - {3 m_t^2 \lambda \over 16 \pi^2 v \sin \beta}
        (A_t + 2 \lambda x \cot \beta)
        f(m_{\tilde{t_1}}^2,m_{\tilde{t_2}}^2)  \ ,
\end{eqnarray}
with
\begin{eqnarray}
        \Delta_1  & = & {m_t^2 \lambda x \over v \sin \beta}
        (A_t + \lambda x \cot \beta)             \cr
        & &\mbox{} + {\cos \beta \over 2 v}
        \left [ (m_Q^2 - m_T^2) + ({4\over 3} m_W^2
        - {5\over 6} m_Z^2) \cos 2 \beta \right]
        ({4\over 3} m_W^2 - {5\over 6} m_Z^2)  \ , \cr
        \Delta_2  &=& {m_t^2 A_t\over v \sin \beta}
        (A_t + \lambda x \cot \beta)             \cr
        & &\mbox{} - {\sin \beta \over 2 v}
        \left [ (m_Q^2 - m_T^2) + ({4\over 3} m_W^2
        - {5\over 6} m_Z^2) \cos 2 \beta \right]
        ({4\over 3} m_W^2 - {5\over 6} m_Z^2)  \ .
\end{eqnarray}
Here two functions $f$ and $g$ are defined as
\begin{eqnarray}
        f(m_1^2,m_2^2) &=& {1 \over (m_2^2-m_1^2)}
        \left[  m_1^2 \log {m_1^2 \over \Lambda^2} -m_2^2
        \log {m_2^2 \over \Lambda^2}
        \right] + 1  \ , \cr
        g(m_1^2,m_2^2) &=& {1 \over (m^2_1 - m^2_2)^3}
        \left [(m_1^2+m_2^2) \log {m_2^2 \over m_1^2}
        - 2 (m_2^2 - m_1^2) \right ] \ ,
\end{eqnarray}
$m_Z^2 = (g_1^2 +g_2^2)v^2/2$ and $m_W^2 = (g_2 v)^2/2$ for
$v = \sqrt{v_1^2 + v_2^2} = $ 175 GeV.
We remark that the mass-squared matrix for the scalar Higgs boson contains
the partial gauge boson contributions.
The analytic formulas for the three scalar Higgs boson masses are given in
terms of $M_{S_{ij}}$ in the previous paper [7].
Here we sort these masses as $m_{S_1} < m_{S_2} < m_{S_3}$.

The mass-squared matrix for the pseudoscalar Higgs boson is given by a
symmetric $3 \times 3$ matrix.
Its mass-squared matrix can be reduced into a symmetric $2 \times 2$ matrix
$M^P$ since there is a massless neutral Goldstone boson.
The elements of the mass-squared matrix for the pseudoscalar Higgs boson up to
the 1-loop level are given as
\begin{eqnarray}
   M_{P_{11}} & = &
      {\lambda x (A_{\lambda} + k x) \over \sin \beta \cos \beta}
      + {3 \over 16 \pi^2 v^2}
      {m_t^2 A_t \lambda x \over \sin^3 \beta \cos \beta}
      f(m_{\tilde{t_1}}^2,m_{\tilde{t_2}}^2) \cr
   M_{P_{22}} & = & \lambda v^2 \sin 2 \beta
      \left (2 k + {A_{\lambda} \over 2 x} \right ) + 3 k A_k x
      + {3 m_t^2 A_t \lambda \cot \beta \over 16 \pi^2 x}
      f(m_{\tilde{t_1}}^2,m_{\tilde{t_2}}^2)    \cr
   M_{P_{12}} & = & \lambda v (A_{\lambda} - 2 k x)
      + {3 m_t^2 \lambda A_t \over 16 \pi^2 v \sin^2 \beta}
      f(m_{\tilde{t_1}}^2,m_{\tilde{t_2}}^2)     \ .
\end{eqnarray}
Note that the mass-squared matrix for the radiatively corrected pseudoscalar
Higgs boson does not changed by the inclusion of terms proportional to gauge
couplings in the top-scalar-quark masses.
Two eigenvalues of the matrix are the mass-squareds of the pseudoscalar Higgs
boson,
\begin{eqnarray}
      m_{P_1,\, P_2}^2 & = & {1 \over 2}
      \left [ Tr M^P \mp \sqrt{(Tr M^P)^2 - 4 det (M^P)} \right ] \ ,
\end{eqnarray}
$m_{P_1} < m_{P_2}$.

At LEP 2 there are three dominant processes for the production of the scalar
Higgs boson.
The most prominent process  for small $\tan \beta$ ($ \approx 2$) in the NMSSM
is the Higgs-strahlung process,
\begin{equation}
\begin{array}{ccccccc}
e^+e^- & \rightarrow & Z & \rightarrow & Z^* S_i
& \rightarrow & {\bar f} f S_i  \ ,
\end{array}
\end{equation}
where the scalar Higgs bosons are emitted from the $Z$ boson.
The next one is the so-called bremsstrahlung process,
\begin{equation}
\begin{array}{ccccccc}
e^+e^- & \rightarrow & Z & \rightarrow & f {\bar f}
& \rightarrow & {\bar f} f S_i   \ ,
\end{array}
\end{equation}
where the pair of fermions comes from the decay of the $Z$ boson.
Then the scalar Higgs boson can be radiated from either of the quark pairs.
It is well known that the bremsstrahlung process for large $\tan \beta$
($ \ge 6$) in the NMSSM gives more dominant contributions to the scalar Higgs
production than the Higgs-strahlung process.
These two processes of the scalar Higgs production are also present in the SM.
However, there is an additional process in the NMSSM, which is equally
important as the above two processes. It is because the NMSSM possesses the
pseudoscalar Higgs bosons $P_j$. The process, called the associated pair
production process, involves $P_j$ as
\begin{equation}
\begin{array}{ccccccc}
e^+e^- & \rightarrow & Z & \rightarrow & P_j h
& \rightarrow & {\bar f} f h  \ ,
\end{array}
\end{equation}
where the pair of fermions comes from the decay of the pseudoscalar Higgs
bosons.
Here the dominant contributions come from bottom quarks, $f = b$,
to the cross section of the scalar Higgs productions via the above three
processes in the NMSSM.

The relevant couplings at the presence of bottom quarks are given by
\begin{eqnarray}
G_{Z Z S_i} & = & {g_2 m_Z \over \cos \theta_W}
\left [ U_{S_{i,\,1}} \cos \beta + U_{S_{i,\,2}} \sin \beta \right ]  \cr
G_{b \bar{b} S_i} & = & {g_2 m_b \over 2 m_W \cos \beta}
\left [ U_{S_{i,\,1}} \right ]                                 \cr
G_{b \bar{b} P_j} & = & {g_2 m_b \over 2 m_W \cos \beta}
\left [ U_{P_{j,\,1}} \right ]                                 \cr
G_{Z S_i P_j} & = & {g_2 \over 2 \cos \theta_W}
\left [ U_{S_{i,\,2}} U_{P_{j,\,2}} - U_{S_{i,\,1}} U_{P_{j,\,1}} \right ] \ ,
\end{eqnarray}
where $U_{S_{i,\,j}}$ ($U_{P_{j,\,i}}$) are the elements of the transformation
matrix which diagonalize the scalar (pseudoscalar) Higgs boson mass-squared
matrix.
Analytic formulas for the elements $U_{S_{i,\,j}}$ are given in terms of
$M_{S_{ij}}$ and $m_{S_i}^2$ in the previous paper [7].
The transformation matrix which diagonalize the pseudoscalar Higgs boson
msss-squared matrix is given by
\begin{eqnarray}
U^P & = &
    \left ( \begin{array}{ccc}
    \cos \alpha \sin \beta   & \cos \alpha \cos \beta    & \sin \alpha  \cr
    - \sin \alpha \sin \beta & - \sin \alpha \cos \beta  & \cos \alpha  \cr
    \cos \beta               & - \sin \beta              &  0
  \end{array} \right ) \ ,
\end{eqnarray}
with
\begin{eqnarray}
\alpha & = &
\tan^{-1} \left (
{\displaystyle {m_{P_1}^2 - M_{P_{11}} \over M_{P_{12}} } }
\right ) \ .
\end{eqnarray}

First let us set a reasonable parameter space in the model in order to
calculate the neutral Higgs boson mass as well as the production cross section
of the scalar Higgs boson.
The CDF and D0 collaborations [8], respectively, predict that the
top-quark mass is $176 \pm 8 \pm 10$ and $199^{+19}_{-21} \pm 22$ GeV.
Therefore we take the top-quark mass of 175 GeV in our numerical calculations.
The upper bounds on $\lambda$ and $k$ are given as 0.87 and 0.63, respectively,
by the renormalization group analysis of the NMSSM [9].
By the renormalization group analysis the lower limits on $\tan \beta$ at
the electroweak scale are about 1.24 for $m_t$ = 175 GeV and 2.6 for
$m_t$ = 190 GeV [10].
We set the range of $\tan \beta$ as $2 \le \tan \beta \le m_t/m_b \approx 40$.
From the values of $\tan \beta$ and $m_t$ = 175 GeV, the value of the
top-quark Yukawa coupling is given by the equation, $h_t = m_t/(v \sin \beta)$.
Then the maximum values of the coupling $\lambda$ are given by $h_t$.
The numerical result is plotted in figures of Ref. [11].
In our numerical analysis $\lambda$ can be a value between zero and its
maximum value for $\tan \beta$ and $m_t$ = 175 GeV.
Additionally we assume that the lower limit on the lighter top-scalar-quark
mass is greater than the top-quark mass.

We plot in Fig. 1a $m_{S_1}^{\rm (max)}$, $m_{S_2}^{\rm (min)}$, and
$m_{S_3}^{\rm (min)}$ at the tree level as a function of $\tan \beta$,
for $0 < k < 0.63$, 100 GeV $\le A_{\lambda} (A_k, x) \le$ 1000 GeV, and
$0 < \lambda < \lambda^{\rm (max)}$.
Here the superscript index (max) means its maximum value while the superscript
index (min) means its minimum value.
We find from Fig. 1a that the upper bound on the tree level mass of the lightest
scalar Higgs boson is about 100 GeV for $\tan \beta$ = 3 in our parameter
setting.
Of course the lightest scalar Higgs boson at the tree level as well as at the
1-loop level has the lower bound of a zero mass.
Since the lower bound on the tree level mass of $S_2$ is about 33 GeV for
$\tan \beta$ = 2, the production of $S_2$ is kinematically allowed at the
proposed center of mass energies of 175, 192, and 205 GeV for LEP 2
via three dominant processes.
The heaviest scalar Higgs boson can be produced at LEP 2 with
$\sqrt{s}$ = 175 GeV because the lower bound on its mass is about 92 GeV
for $\tan \beta$ = 16.
Nevertheless we surmise that the production cross section of $S_3$ is very
small at three c.m energies for LEP 2.
Thus we scan for the parameter space of $0 < k < 0.63$,
100 GeV $\le A_{\lambda} (A_k, x) \le$ 1000 GeV, and
$0 < \lambda < \lambda^{\rm (max)}$ and plot the lower bounds on
($\sigma_{S_1} + \sigma_{S_2}$) at the tree level as a function of $\tan \beta$
at three c.m energies in Fig. 1b.
The discovery limits in the model are about 50 fb for $m_S$ = 80 GeV and 30 fb
for $m_S$ = 40 GeV at a luminosity of 500 pb$^{-1}$ for $\sqrt{s}$ = 175 GeV
and at one of 300 pb$^{-1}$ for $\sqrt{s}$ = 192 and 205 GeV [12].
We cannot impose any experimental constraints on $\tan \beta$ at $\sqrt{s}$
= 175 GeV.
On the other hand we can impose an experimental constraint on $\tan \beta$
at $\sqrt{s}$ = 192 GeV.
The experimental upper limit on $\tan \beta$ at $\sqrt{s}$ = 192 GeV with the
discovery limit of 50 fb can be set as
\[
\tan \beta_{\rm (EXP)} \le 32.
\]
For LEP 2 with $\sqrt{s}$ = 205 GeV the lower bound on
($\sigma_{S_1} + \sigma_{S_2}$) always is greater than 55 fb in the whole
parameter space.
Thus we find that the Higgs search at LEP 2 with $\sqrt{s}$ = 205 GeV can
cover fully the tree level parameter space of the NMSSM.

Next let us investigate whether the Higgs searches at LEP 2 can give any
experimental constraints on the 1-loop level parameter space of the model.
We fix as $\Lambda$ = $m_Q$ = $m_T$ = $A_t$ = 1000 GeV for the newly introduced
parameters at the 1-loop level and plot in Fig. 2 $m_{S_1}^{\rm (max)}$,
$m_{S_2}^{\rm (min)}$, and $m_{S_3}^{\rm (min)}$ as a function of $\tan \beta$,
for $0 < k < 0.63$, 100 GeV $\le A_{\lambda} (A_k, x) \le$ 1000 GeV, and
$0 < \lambda < \lambda^{\rm (max)}$.
We find from Fig. 2 that the upper bound on the tree level mass of the
lightest scalar Higgs boson can be increased by about 44 GeV in favor of
radiative corrections due to the top-quark and top-scalar-quark loops.
We also find from Fig. 2 that the productions of three scalar Higgs bosons are
kinematically allowed at three c.m energies for LEP 2.
Similarly from the tree level, $S_3$ at LEP 2 c.m energies can be
produced by a small size in its production cross section.
Thus we have calculated ($\sigma_{S_1} + \sigma_{S_2}$) at three c.m energies
for LEP 2 for the parameter space of $2 \le \tan \beta \le 40$,
$0 < k < 0.63$, 100 GeV $\le A_{\lambda} (A_k, x) \le$ 1000 GeV,
$0 < \lambda < \lambda^{\rm (max)}$, and
$\Lambda$ = $m_Q$ = $m_T$ = $A_t$ = 1000 GeV.
We have found that the production cross section of the scalar Higgs boson
at the tree level is fatally modified by radiative corrections to
both the neutral Higgs boson masses and the relevent couplings.
We display their numerical results in Table. 1.
A small fraction of the 1-loop level parameter space of the model will be
investigated by the Higgs search at LEP 2 with a discovery limit of
30 $\sim$ 50 fb.

%***********************************************************************
\section{Conclusions}
%***********************************************************************

\hspace*{5.5mm}
We have investigated the neutral Higgs sector of the NMSSM at LEP 2.
We have calculated the lower bounds on the scalar Higgs boson masses at
the tree level as well as at the 1-loop level.
Especially, the scalar Higgs boson masses at the 1-loop level are changed by
the inclusion of terms proportional to gauge couplings in the top-scalar-quark
mass.
Thus these bounds at the 1-loop level contain the partial gauge boson
contributions.
At the tree level the production cross section ($\sigma_{S_1} + \sigma_{S_2}$)
is calculated at three c.m energies for LEP 2 via three dominant processes.
We cannot put any experimental constraints on $\tan \beta$ at LEP 2 with
$\sqrt{s}$ = 175 GeV.
On the other hand an experimental upper limit on $\tan \beta$ at the tree level
can be set by $\tan \beta_{\rm (EXP)} \le$ 32 for LEP 2 with $\sqrt{s}$ = 192
GeV.
We also find that the Higgs search at LEP 2 with $\sqrt{s}$ = 205 GeV is
sufficient to cover the whole parameter space of the model at the tree level.
The production cross section of the scalar Higgs boson at the tree level is
significantly affected by radiative corrections both to the neutral Higgs boson
mass and to the relevant couplings.
We expect that the 1-loop level parameter space of the model at LEP 2 can
be explored 6.9 percent for $\sqrt{s}$ = 175 GeV, 8.0 percent for
$\sqrt{s}$ = 192 GeV, and 8.4 percent for $\sqrt{s}$ = 205 GeV.

%***********************************************************************

%*********************************************************************
\vfil \eject
%*********************************************************************

{\bf Figure Captions}
\vskip 0.3 in
\noindent
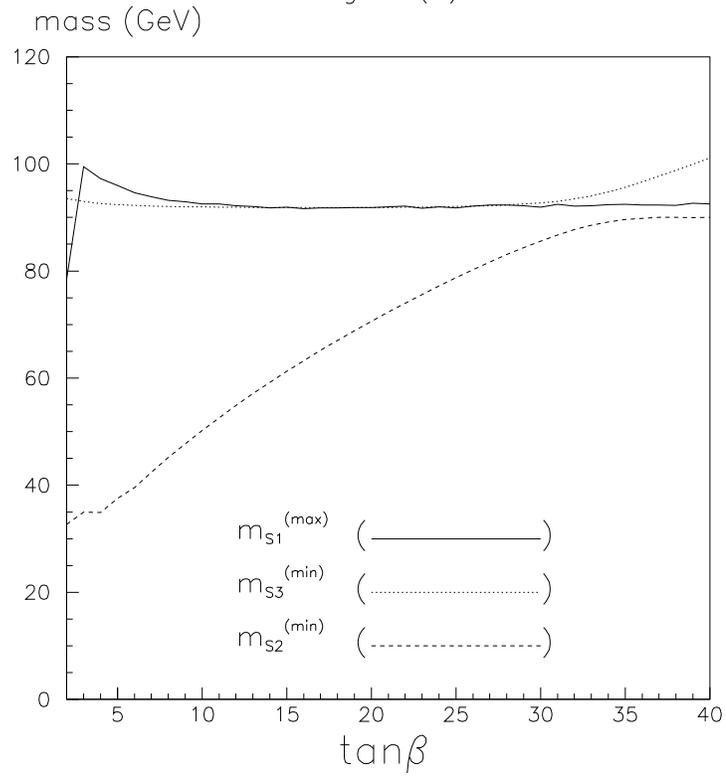
Fig. 1. \  (a) $m_{S_1}^{\rm (max)}$, $m_{S_2}^{\rm (min)}$, and
$m_{S_3}^{\rm (min)}$ (b) the upper bounds of the cross section
($\sigma_{S_1} + \sigma_{S_2}$) at three c.m energies, as a function of
$\tan \beta$, for $0 < k < 0.63$,
100 GeV $\le A_{\lambda} (A_k, x) \le$ 1000 GeV, and
$0 < \lambda < \lambda^{\rm (max)}$ at the tree level.
\vskip 0.2 in
\noindent
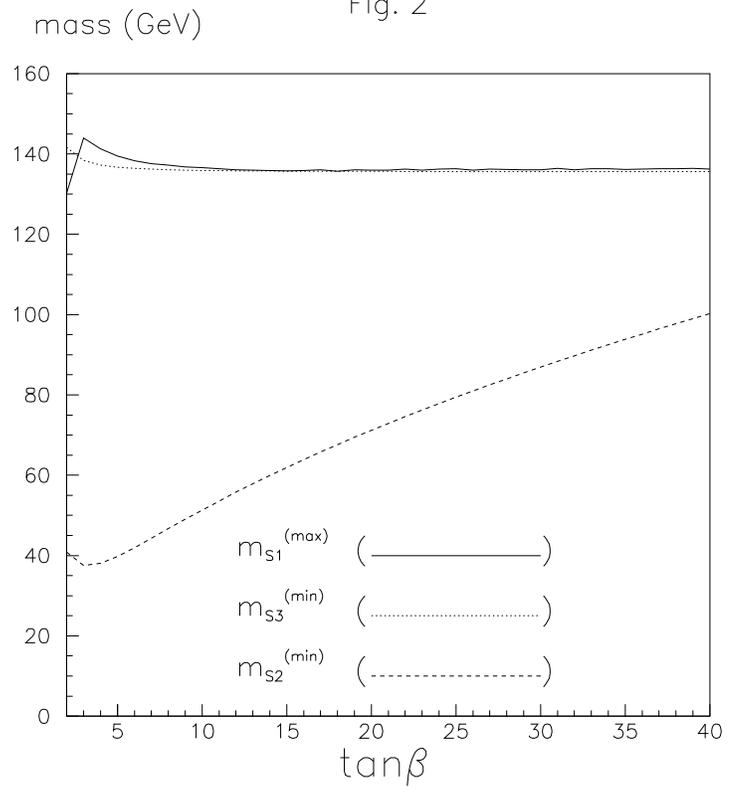
Fig. 2. \  $m_{S_1}^{\rm (max)}$, $m_{S_2}^{\rm (min)}$, and
$m_{S_3}^{\rm (min)}$ at the 1-loop level as a function of $\tan \beta$,
for $0 < k < 0.63$, 100 GeV $\le A_{\lambda} (A_k, x) \le$ 1000 GeV,
$0 < \lambda < \lambda^{\rm (max)}$, and
$\Lambda$ = $m_Q$ = $m_T$ = $A_t$ = 1000 GeV.
\vskip 0.2 in
\vfil\eject
\noindent
%*****************************************************************************

\begin{center}
Table 1. The numerical results for $(\sigma_{S_1} + \sigma_{S_2})$ at three c.m
energies for the parameter space of 2 $\le$ $\tan \beta$ $\le$ 40,
0 $< k <$ 0.63, 100 GeV $\le$ $A_{\lambda}$ ($A_k, x$) $\le$ 1000 GeV,
0 $<$ $\lambda$ $<$ $\lambda^{\rm (max)}$, and
$\Lambda$ = $m_Q$ = $m_T$ = $A_t$ = 1000 GeV.
The allowed point number for \ \ ($\sigma_{S_1}$ + $\sigma_{S_2}$) in the parameter
space is about 29756 points at three c.m energies for LEP 2.

\begin{tabular} {c|c|c|c}
\hline \hline
$\sqrt{s}$    & $(\sigma_{S_1} + \sigma_{S_2}) \ge$ 50 (fb)  &
$(\sigma_{S_1} + \sigma_{S_2})^{\rm (max)}$  &
$(\sigma_{S_1} + \sigma_{S_2})^{\rm (min)}$
\cr
\hline
175 (GeV)     & 2050 (points) & 700 (fb)  & 0.005 (fb) \cr
192 (GeV)     & 2380 (points) & 900 (fb)  & 0.03 (fb) \cr
205 (GeV)     & 2490 (points) & 1010 (fb) & 0.09 (fb) \cr
\hline \hline
\end{tabular}
\end{center}
\vfil\eject
%***********************************************************************
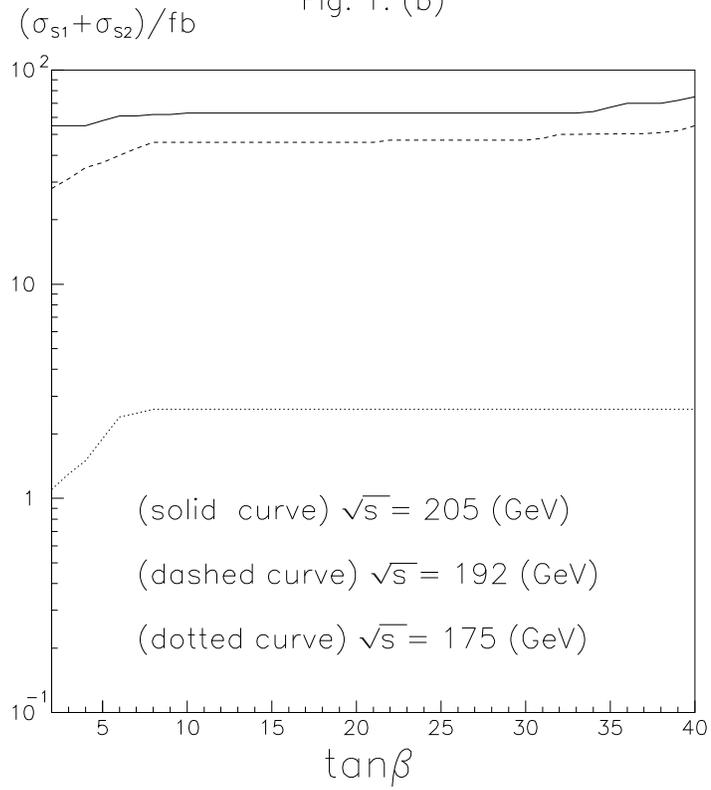
\end{document}